\documentclass[sigconf]{acmart}
\usepackage{algorithmic}
\usepackage{graphicx}
\usepackage{textcomp}
\usepackage{graphicx}
\usepackage{caption}
\usepackage{subcaption}
\usepackage{bbm}
\usepackage{color}
\usepackage{hyperref}
\usepackage{multirow}
\usepackage{enumitem}
\usepackage{algorithm} 
\usepackage{algorithmic}
\usepackage{enumerate}
\usepackage{amsmath}
\usepackage{booktabs}
\usepackage{multirow}
\usepackage{stmaryrd}
\usepackage{mathrsfs} 
\usepackage{color}
\usepackage{balance}
\usepackage{adjustbox}

\usepackage[para,online,flushleft]{threeparttable}

\newcommand{\eg}{\textit{e.g.}}

\newcommand{\ie}{\emph{i.e.}}

\setlist[itemize]{leftmargin=*}
\renewcommand\footnotetextcopyrightpermission[1]{}
\AtBeginDocument{%
  \providecommand\BibTeX{{%
    \normalfont B\kern-0.5em{\scshape i\kern-0.25em b}\kern-0.8em\TeX}}}

\setcopyright{acmlicensed}
\copyrightyear{2018}
\acmYear{2018}
\acmDOI{XXXXXXX.XXXXXXX}

\acmConference[Conference acronym 'XX]{Make sure to enter the correct
  conference title from your rights confirmation emai}{June 03--05,
  2018}{Woodstock, NY}
\acmBooktitle{Woodstock '18: ACM Symposium on Neural Gaze Detection,
 June 03--05, 2018, Woodstock, NY} 
\acmISBN{978-1-4503-XXXX-X/18/06}

\begin{document}

\title{MemoCRS: Memory-enhanced Sequential Conversational Recommender Systems with Large Language Models}


\author{Yunjia Xi}
\email{xiyunjia@sjtu.edu.cn}
\affiliation{%
  \institution{Shanghai Jiao Tong University}
  \city{Shanghai}
  \country{China}
}

\author{Weiwen Liu}
\email{liuweiwen8@huawei.com}
\affiliation{%
  \institution{Huawei Noah's Ark Lab}
  \city{Shenzhen}
  \country{China}
}

\author{Jianghao Lin}
\email{chiangel@sjtu.edu.cn}
\affiliation{%
  \institution{Shanghai Jiao Tong University}
  \city{Shanghai}
  \country{China}
}

\author{Bo Chen}
\email{chenbo116@huawei.com}
\affiliation{%
  \institution{Huawei Noah's Ark Lab}
  \city{Shenzhen}
  \country{China}
}

\author{Ruiming Tang}
\email{tangruiming@huawei.com}
\affiliation{%
  \institution{Huawei Noah's Ark Lab}
  \city{Shenzhen}
  \country{China}
}

\author{Weinan Zhang}
\email{wnzhang@sjtu.edu.cn}
\affiliation{%
  \institution{Shanghai Jiao Tong University}
  \city{Shanghai}
  \country{China}
}

\author{Yong Yu}
\email{yyu@sjtu.edu.cn}
\affiliation{%
  \institution{Shanghai Jiao Tong University}
  \city{Shanghai}
  \country{China}
}

\renewcommand{\shortauthors}{Yunjia Xi et al.}


\begin{abstract}
  Conversational recommender systems (CRSs) aim to capture user preferences and provide personalized recommendations through multi-round natural language dialogues. However, most existing CRS models mainly focus on dialogue comprehension and preferences mining from the current dialogue session, overlooking user preferences in historical dialogue sessions. The preferences embedded in the user's historical dialogue sessions and the current session exhibit continuity and sequentiality, and we refer to CRSs with this characteristic as \textit{sequential CRSs}. In this work, we leverage memory-enhanced LLMs to model the \textit{preference continuity}, primarily focusing on addressing two key issues: (1) redundancy and noise in historical dialogue sessions, and (2) the cold-start users problem. To this end, we propose a \underline{Memo}ry-enhanced \underline{C}onversational \underline{R}ecommender \underline{S}ystem Framework with Large Language Models (dubbed \textit{MemoCRS}), consisting of user-specific memory and general memory. User-specific memory is tailored to each user for their personalized interests and implemented by an \textit{entity-based memory bank} to refine preferences and retrieve relevant memory, thereby reducing the redundancy and noise of historical sessions. The general memory, encapsulating \textit{collaborative knowledge} and \textit{reasoning guidelines}, can provide shared knowledge for users, especially cold-start users. With the two kinds of memory, LLMs are empowered to deliver more precise and tailored recommendations for each user. Extensive experiments on both Chinese and English datasets demonstrate the effectiveness of MemoCRS. 
\end{abstract}

\maketitle
\section{Introduction}

%

\begin{figure}
    \centering
    \vspace{-5pt}
    \includegraphics[trim={0.9cm 0.3cm
    0 0},clip,width=0.51\textwidth]{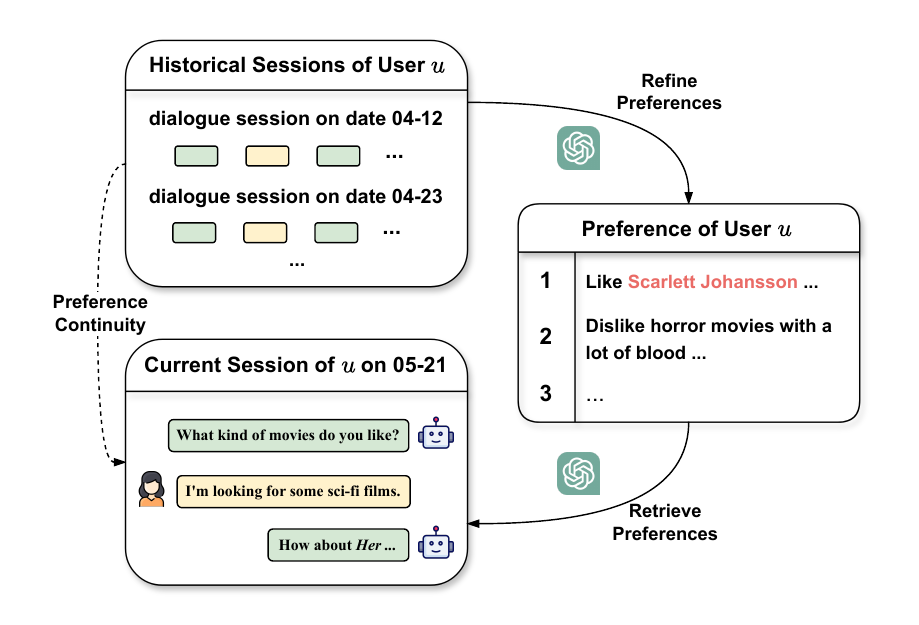}
    \vspace{-20pt}
    \caption{An example of leveraging the user preference continuity to assist in conversational recommendations. The yellow and green rectangles denote the utterances from the user and the system, respectively. }
    \vspace{-10pt}
    \label{fig:intro}
\end{figure}

Conversational recommender systems (CRSs) engage users in multi-turn dialogue, aiming to elicit user preferences and provide personalized recommendations~\cite{jannach2021survey,tgredial,redial,crslab,sun2018conversational}. Unlike traditional recommendation systems that rely solely on user-item interactions~\cite{liu2022neural,zhang2021deep,lin2023map}, \eg, clicks and purchases, CRSs can comprehend natural language instructions, gather users' real-time feedback, and figure out user preferences from ongoing conversations. Hence, it is expected to recommend items that exactly match the user's need with human-like responses. To this end, CRSs typically comprise two essential components: a recommender to provide recommendations aligned with user preferences, and a generator to produce natural language responses~\cite{uccr,UniCRS,zesrec,UniSRec}. 
The completion of both components' functionalities hinges on the system's ability to comprehend dialogue nuances and uncover user preferences accurately.

However, the predominant focus of most CRSs is mainly on dialogue comprehension and preferences mining from the \textit{current dialogue session}, neglecting user preferences reflected in \textit{historical dialogue sessions}. Here, a dialogue session refers to a complete multi-turn conversation over a continuous period of time, starting from the user or the system initiating the conversation until the user ends the conversation and leaves~\cite{uccr}. As the dialogue serves as the primary medium for communication and recommendation in the conversational recommendation, traditional CRSs typically prioritize enhancing the dialogue comprehension~\cite{uccr} to extract user interests more effectively, such as employing more complex encoders~\cite{UniCRS,kgsf} or leveraging external information like knowledge graphs (KGs)~\cite{kbrd,kgsf,zhang2023variational} and reviews~\cite{yang2021improving,lu2021revcore}. However, they tend to overlook the central actors in CRSs, \ie, \textbf{users}, whose behaviors and preferences exhibit continuity across sessions. Actually, preference continuity---the consistent patterns and tendencies in users' behaviors and preferences over time---is critical for recommendation accuracy \cite{uccr,sun2024large}. This leads CRSs to exhibit sequentiality and coherence, akin to sequential recommendations~\cite{kang2018self,gru4rec,sun2019bert4rec,liu2024mamba4rec}, which we can refer to as \textit{sequential CRSs}. In sequential CRSs, users may display varying preferences across sessions, which coexist and interrelate with each other. Incorporating these historical preferences can help us better comprehend the current dialogue session and uncover some nuanced and implicit interests~\cite{sun2024large}. For instance, in Figure~\ref{fig:intro}, if the user's historical sessions reveal a preference for the actress \textit{Scarlett Johansson}, and now the user asks for sci-fi movies, then sci-fi movies starring \textit{Scarlett Johansson}, like ``\textit{Her}'', would likely align well with the user's demand and preference. 

Recently, large language models (LLMs) such as ChatGPT~\cite{gpt4} have showcased remarkable proficiency in comprehending and generating natural languages~\cite{qiao2023reasoning,zhao2023survey,zhu2023large}. 
While there are several explorations of applying LLMs to CRSs, yet none of these works touches upon modeling the user's preference continuity through historical dialogues. 
They usually focus merely on zero-shot recommendations on the current session~\cite{zesrec,wang2023rethinking}, evaluation of CRSs as user simulators~\cite{wang2023rethinking,yoon2024evaluating,zhu2024reliable,yang2024behavior}, or data augmentation for CRSs~\cite{wang2023improving}.
Nevertheless, in the realm of conversational agents, researchers leverage the updatable textual memory as a plug-in unit for LLMs to handle long-range dependencies and contexts~\cite{modarressi2023ret,zhao2024expel,liu2023think}, as well as modeling user personalities and preferences effectively~\cite{zhong2024memorybank,packer2023memgpt,lu2023memochat}.
Therefore, we, for the first time, propose to bring memory-enhanced LLMs to sequential CRSs, allowing for the modeling of user preference continuity with updatable textual memory mechanisms.

 
Although the updatable textual memory is a popular choice for plug-and-play, interpretable, transparent, and scalable memory mechanisms of LLMs, it generally suffers from the following two key challenges for LLM-based sequential CRSs. 
 \textit{Firstly}, the user's historical dialogue sessions usually contain redundant, irrelevant, and noisy information, making it suboptimal to simply build a memory bank over all historical dialogues.
 As shown in Figure~\ref{fig:intro}, historical sessions can be refined by LLMs into concise preferences knowledge to remove redundancy. Not all the preferences are relevant to the ongoing current conversation where the user is seeking a sci-fi movie. For example, the preference for horror movies may introduce noise as it is irrelevant to the user's current demand.
Therefore, it is imperative to refine preferences and retrieve relevant ones. 
\textit{Secondly}, the recommendation depends not only on the user's individual preferences but also on the universal knowledge shared among users, \eg, collaborative knowledge. 
This knowledge necessitates a holistic understanding of data distribution, an aspect LLMs often grapple with. Furthermore, not all users have sufficient historical conversations to establish personalized memory, leading to the problem of cold-start users with limited memory~\cite{lam2008addressing}. Thus, it is also crucial to preserve general knowledge shared among users.

To tackle the above problems, we propose a \underline{Memo}ry-enhanced \underline{C}onversational \underline{R}ecommender \underline{S}ystem Framework with Large Language Models (dubbed \textbf{MemoCRS}) to capture the user preference continuity for sequential CRSs. 
Specifically, we devise two types of textual memory: (1) user-specific memory and (2) general memory. User-specific memory is tailored to each user for individual and personalized preferences. We implement this through an \textit{entity-based memory bank}, housing entities like item names and attributes mentioned in historical dialogues alongside the associated user attitudes and timestamps. This structured memory bank supports operations such as \textit{add}, \textit{merge}, \textit{retrieve}, and \textit{delete}. When a new dialogue occurs, LLMs retrieve relevant memories from this bank to assist in recommendations, thereby mitigating redundancy and noise.  General memory contains shared and universal knowledge among users that transcends individual dialogues. In our framework, we primarily focus on two core aspects: \textit{collaborative knowledge}, which contains shared preference patterns among different users, and \textit{reasoning guidelines} that guide the reasoning process of LLMs. The former is provided by external expert models, while the latter is self-reflectively summarized by LLMs. LLMs can leverage this external and self-summarized knowledge to provide users, especially cold-start users, with suitable recommendations. Finally, incorporating those memories and the current conversation, LLMs are empowered to deliver more precise and tailored recommendations for each user. Our main contributions can be summarized as follows:
\begin{itemize}
    \item We emphasize the pivotal role of user preference continuity in sequential CRSs and leverage the textual memory to extract and store user preferences. To the best of our knowledge, this is the first work to explicitly use memory-enhance LLMs to refine and manage user preferences in sequential CRSs. 
    \item We propose a \underline{Memo}ry-enhanced \underline{C}onversational \underline{R}ecommender \underline{S}ystem Framework with Large Language Models (dubbed \textbf{MemoCRS}), where two types of memory are devised: user-specific memory for users' personalized preferences and general memory for shared experience and cold-start users,  both of which significantly enhance the modeling of user preference continuity.
    \item The textual memory mechanism in MemoCRS is highly plug-and-play, interpretable, transparent, and scalable for LLMs, ensuring efficient and effective recommendations in sequential CRSs.
\end{itemize}
\vspace{-5pt}
Extensive experiments on both Chinese and English datasets demonstrate that MemoCRS significantly outperforms traditional and LLM-based baselines. We believe MemoCRS sheds light on a way to model the user preference continuity in sequential CRSs.

\vspace{-5pt}
\section{Related Work}
\subsection{Conversational Recommender Systems}
CRSs aim to capture user preferences and provide personalized recommendations through multi-round natural language dialogues~\cite{gao2021advances}. Based on how the response is generated, CRSs can be divided into two categories: attribute-based and generation-based models~\cite{feng2023large}. The former uses pre-defined actions (\eg, asking queries about item attributes and generating responses with pre-defined templates) to interact with users~\cite{lei2020estimation,christakopoulou2016towards,he2022bundle,lei2020estimation}. They aim to capture user interests and provide recommendations that users accept in as few rounds as possible~\cite{crslab,zhang2022multiple}. Conversely, the latter group aims to generate natural and fluent dialogues while delivering high-quality recommendations~\cite{tgredial,redial}. Therefore, models in this line typically have a recommender to generate recommendations and a generator to produce free-form responses. They often utilize knowledge graphs to augment recommender~\cite{kbrd,kgsf,zhang2023variational,UniCRS,yang2021improving} and leverage Pre-trained Language Models (PLMs) as generators for more natural responses~\cite{zhang2023variational,UniCRS,wang2021recindial,yang2021improving}. Early CRSs often used relatively small PLMs, such as DialoGPT for UniCRS~\cite{UniCRS} and GPT-2 for MESE~\cite{yang2021improving}.

The emergence of large language models (LLMs) has brought tremendous success in Natural Language Processing (NLP), and it also shows great potential in other domains like recommendations~\cite{lin2023can,wang2023flip,xi2023towards,fan2023recommender,chen2023large,wu2023survey,clickprompt}. 
Recently, some explorations have been conducted to apply LLMs to CRSs and reveal that LLMs exhibit a deep understanding of dialogues and can provide more natural responses and precise recommendations~\cite{zesrec,wang2023rethinking,chatrec}. 
However, current research predominantly revolves around zero-shot recommendations within current session~\cite{zesrec,wang2023rethinking}, evaluation of CRSs such as user simulators~\cite{wang2023rethinking,yoon2024evaluating,zhu2024reliable,yang2024behavior}, data augmentation for CRSs~\cite{wang2023improving}, and sub-task management and planning~\cite{feng2023large,li2023long}. 
Those works neglect the important role of user preferences and their continuity reflected in users' historical sessions. Some works involving agents involve user memory units~\cite{huang2023recommender,friedman2023leveraging}, but they merely propose conceptual designs~\cite{friedman2023leveraging} or utilize historical items without considering how to better manage and leverage memory~\cite{huang2023recommender}, like redundancy elimination, noise reduction, and general memory management. To the best of our knowledge, we are the first to introduce memory-enhanced LLMs to refine and manage user preferences in the sequential CRSs, enhancing the performance of the recommendation.

\vspace{-5pt}
\subsection{Memory}
In the development of artificial intelligence, imbuing models with human-like memory has always been a significant research direction. Early studies focused on using parameterized memory in the form of external memory networks~\cite{graves2014neural,graves2016hybrid,weston2014memory,sukhbaatar2015end}. They utilize a memory matrix to store historical hidden states, allowing for effective reading and updating of this matrix. While this approach is convenient, its design is overly simplistic and lacks interpretability and scalability. With the rise of LLMs, especially LLM-based agents~\cite{park2023generative,zhang2024large}, memory has become a crucial component supporting agent-environment interactions. To facilitate input into LLMs, memory is often in natural language form, which enhances interpretability and transparency~\cite{zhong2024memorybank,lu2023memochat,zhao2024expel,liu2023think,modarressi2023ret}. For instance, ExpeL~\cite{zhao2024expel} gathers experience and knowledge using natural language from a collection of training tasks. MoT~\cite{li2023mot} pre-thinks on the unlabeled dataset and saves the high-confidence thoughts as external memory. MemoryBank~\cite{zhong2024memorybank} summarizes relevant memories from previous interactions, evolving through continuous memory updates to adapt to a user's personality.

\begin{figure*}
    \centering
    \vspace{-10pt}
    \includegraphics[trim={0.2cm 0.2cm 0.2cm 0.2cm},clip,width=\textwidth]{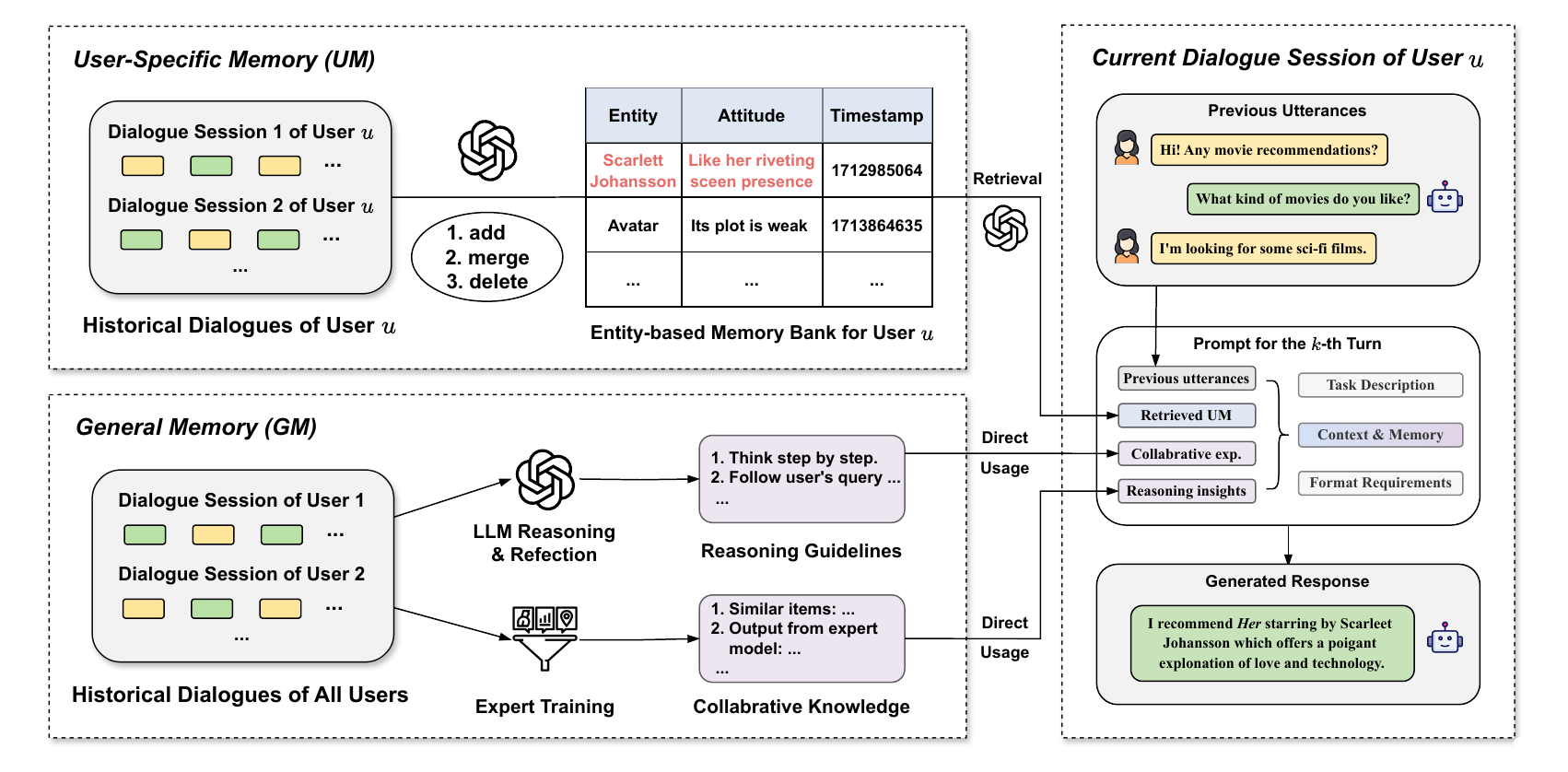}
    \vspace{-20pt}
    \caption{The overall framework of MemoCRS.}
    \vspace{-10pt}
    \label{fig:framework}
\end{figure*}
\vspace{-5pt}
\section{Preliminaries}\label{sect:preliminaries}
Conversational recommender systems (CRSs) aim to elicit user preferences and provide precise item recommendations through multi-turn natural language dialogues. Therefore, CRSs typically consist of a recommender to generate recommendations that match user preferences and a generator to produce natural language responses based on the recommendation~\cite{uccr,UniCRS,tgredial,kbrd,kgsf}. To achieve this, both the recommender and generator rely on comprehending the dialogue and uncovering user preferences. Previous works primarily focus on dialogue comprehension and preference mining for the user's current dialogue session, overlooking the user's preferences within their historical sessions. In this work, we propose that in the real world, a user may engage in multiple dialogue sessions with CRSs, where the user's historical dialogues and preferences exhibit sequentiality and continuity. We refer to this as Sequential CRSs and present its specific formulation as follows.

Formally, let $\mathcal{U}$, $\mathcal{I}$, and $\mathcal{V}$ denote the user set, item set, and the vocabulary. For a user $u\in\mathcal{U}$ who has $T$ conversation sessions with the system, we organize all his/her dialogue sessions in chronological order and refer to them as $\{C_t\}^T_{t=1}$.  Each dialogue $C_t$ consists of $n$ utterances denoted as $C_t=\{s_k\}_{k=1}^n$, where $s_k$ represents the utterance at the $k$-th turn and each utterance $s_k$ is composed by $m$ words from vocabulary $\mathcal{V}$, \ie, $s_k=\{w_j\}^m_{j=1}$. 
In each utterance $s_k$, the user or recommender may mention some items denoting $\mathcal{I}_k\in\mathcal{I}$. 
We regard the last session $C_T$ as the \textit{current dialogue session} in which we aim to provide recommendations. Then, all the sessions before the current dialogue session are denoted as \textit{historical dialogue sessions} $H_u = \{C_t\}_{t=1}^{T-1}$ for the user $u$. 

It is worth noting that, following previous work~\cite{uccr}, our definition of \textit{historical conversation/dialogue sessions} differs from the \textit{conversation/dialogue history} commonly used in most CRSs works~\cite{kbrd,kgsf,UniCRS}.   Conversation/dialogue history is historical utterances preceding the current utterance within the same session, pertaining to turn-level. Historical dialogue sessions refer to complete dialogues occurring at different times before the current session.

Based on the above definition, the task of sequential conversational recommendation can be defined as follows. At the $k$-th turn (the turn the recommender should speak), given the historical dialogue sessions $H_u=\{C_t\}^{T-1}_{t=1}$ and the current dialogue session before $k$-th turn, \ie, $\{s_j\}_{j=1}^{k-1}$, CRSs need to select a candidate set $\hat{\mathcal{I}_k}$ from the entire item candidate set $\mathcal{I}$, such that $\hat{\mathcal{I}_k}$ is as close to the user's current needs and implicit preferences as possible, and generate reasonable responses for the items in $\hat{\mathcal{I}_k}$.

 

\section{Methodology}

\subsection{Overview of MemoCRS}
This framework of MemoCRS, as shown in Figure~\ref{fig:framework}, is model-agnostic and consists of two types of memory: (1) user-specific memory for user's personalized preferences, and (2) general memory for shared and universal knowledge among users.

\smallskip
\noindent
\textbf{User-Specific Memory (UM)} is a unique \textit{entity-based memory bank} for each user, designed to store the user's individual and personalized preferences. It is a dynamic memory bank that stores the entities mentioned in the user's historical dialogues (\eg, item names and attributes), along with the associated user attitudes and timestamps. It is also scalable and updatable, supporting multiple memory operations, such as \textit{add}, \textit{merge}, \textit{retrieve}, and \textit{delete}.

\smallskip
\noindent
\textbf{General Memory (GM)} preserves shared and universal knowledge among diverse users that cannot be derived from individual dialogues alone. It is primarily composed of \textit{collaborative knowledge} related to recommendations and \textit{reasoning guidelines} for the reasoning of LLMs. These two types of knowledge maintain their compactness, enabling direct utilization without memory retrieval. The universality of the knowledge enables LLMs to infer the preferences of cold-start users with limited historical dialogues, thus enhancing recommendations.

When dealing with conversational recommendations, the general memory and relevant entities and attitudes retrieved from user-specific memory are incorporated into the prompt alongside the current conversation context. This integration enables LLMs to generate recommendations aligned with the user's immediate needs and implicit preferences.

\subsection{User-Specific Memory (UM)}\label{sec:PM}
While users exhibit different preferences across various conversations, these preferences are continuous and coexist simultaneously due to the consistency of user behaviors. This \textit{preference continuity} describes the consistent patterns and tendencies in users' behaviors and preferences over time, which is the key to recommendation tasks~\cite{uccr,sun2024large}. Understanding the preferences manifested in a user's historical sessions can aid in uncovering their implicit needs in the current conversation, facilitating recommendations that are more aligned with their requirements. 
For example, if a user's historical dialogues reveal a preference for the actress \textit{Scarlett Johansson}, and now the user asks for some sci-fi movies, then a sci-fi movie starring \textit{Scarlett Johansson}, like \textit{Her}, may be an excellent choice. 
Inspired by recent studies on memory~\cite{lu2023memochat,zhong2024memorybank,modarressi2023ret,zhao2024expel}, we propose to leverage an external memory bank to record each user's historical preferences, thereby assisting LLMs in more personalized recommendations. 

However, simply preserving all the historical dialogues or items as memory is not practical\cite{uccr,huang2023recommender}. We point out that since user preferences may be multifaceted and the user often has specific needs in the current conversation, not all historical sessions or items will be helpful for the current conversation. 
This approach may introduce unnecessary redundancy and noise. 
Moreover, when users have extensive historical dialogues, including all of them may exceed the context windows of LLMs.
Some researchers also find that LLMs often fail to extract useful information for the recommendation from a textual context of long user behavior sequence, even if the length of context is far from
reaching the context limitation of LLMs~\cite{lin2023rella}. Therefore, we need to compress and refine the user's historical dialogue sessions and only extract memories relevant to the user's ongoing conversational demands.

Thus, we have devised an entity-based memory bank, which comprises three kinds of crucial information: \textit{entity}, user's \textit{attitudes} towards the entity, and \textit{timestamp}. LLMs extract entities mentioned in users' historical sessions, such as movie titles, actors, directors, and genres for movie scenes, along with users' attitudes toward them. Moreover, this memory bank is dynamic and expandable, supporting operations like \textit{add}, \textit{merge}, \textit{retrieve}, and \textit{delete}.

\subsubsection{Memory Storage}
In user-specific memory, each user $u$ necessitates a separate memory bank $\mathcal{M}_u$ dedicated to storing their preferences. We extract pivotal information pertinent to user preferences from historical dialogue sessions: entities mentioned by the user, the nuanced attitudes the user holds towards these entities, and timestamp. Here, we formulate memory bank $\mathcal{M}_u$ as a dictionary, where entity serves as key, and others act as values, \ie, 

\begin{equation}
    \mathcal{M}_u = \left\{\,\text{Entity}_i \rightarrow \big(\text{Attitude}_i,\text{Timestamp}_i\big)\,\right\}_{i=1}^{|\mathcal{M}_u|}, \; u \in \mathcal{U}.
\end{equation}

Since the number of entities mentioned by users is limited, knowledge extracted in this way is more compact and more relevant to recommendations than the whole dialogue. This approach not only removes redundancy to improve storage efficiency but also facilitates updates and retrieval, as we can find relevant attitudes based on entities. Moreover, this design draws inspiration from the intricacies of human memory, which typically involves selective compression and summarization rather than maintaining every detail~\cite{baddeley1997human,bates2020efficient}. As humans reminisce, specific keywords emerge as vital cues, aiding in recalling specific narratives, thoughts, and feelings.

\smallskip
\noindent
\textbf{Entity} serves as the key in the memory bank, including the items, attributes, and characteristics of items mentioned by users in historical conversation sessions. Within the domain of recommendation, user interests are primarily reflected in their preferences for items and the attributes of those items. This also constitutes the main content that personalized memory needs to be retained and serves as vital clues guiding our retrieval for specific user preferences. 

\smallskip
\noindent
\textbf{Attitude} refers to a user's specific views and stances towards a particular entity. Retaining just the entity in the memory bank is not sufficient, as a user's preference for a particular entity can be quite complex -- it could be liking or disliking, or it might involve nuanced conditional assessments, \ie, the user likes to watch \textit{Love Actually} during Christmas. For instance, regarding horror movies, a user may dislike those with excessive gore but enjoy psychologically thrilling ones. Moreover, user preferences are dynamic and can evolve over time. Thus, beyond the entity, we need to capture the user's specific and up-to-date attitude toward it. 

\smallskip
\noindent
\textbf{Timestamp} for each entry is the last thing we need to record, denoting the latest moment of operation (comprising add, merge, and retrieval) conducted on each entry. This approach is aimed at facilitating subsequent deletion operations. In situations where storage capacity becomes a concern, we prioritize deletion based on these timestamps, ensuring efficient memory management.

\subsubsection{Memory Update}
As users' dialogue sessions accumulate, their preferences may also undergo constant changes, with new preferences emerging and previous ones potentially being overturned.  Consequently, we must continually update our memory bank $\mathcal{M}_u$ of user $u$ as new dialogue sessions come up. To achieve this, we have devised several operations for updating the memory bank, encompassing add, merge, and delete. 
Apart from the delete operation, the add and merge are all implemented by LLMs. Those prompts, including those used by LLMs thereafter, all consist of three components: \textit{task description}, \textit{context}, and \textit{format requirements}. The task description provides a comprehensive overview of the instructions for the entire task. The context includes inputs that need to be incorporated into the prompt, such as memory and previous utterances of the current conversation. The format requirements specify certain expectations for the output, \eg, \textit{JSON or list} format, facilitating results extraction with a post-hoc text parser. Such a prompt design is flexible and can be adapted to various tasks. 

\textbf{Add} operation is a primary component of memory updating, involving the extraction of entities and attitudes from dialogues and their inclusion in the memory bank $\mathcal{M}_u$. While traditional entity extraction models~\cite{al2020named,nasar2021named} can proficiently handle entity extraction tasks, our work extends to further extracting user attitudes corresponding to these entities, so we adopt LLMs. For a user $u$, upon his/her completion of each conversation session $C_t$, we incorporate the dialogue into a prompt $P_{add}$ and feed it into LLMs as follows,
\begin{equation}
    \{(e_i, a_i)\}^L_{i=1} = f_{LLM}(C_t, P_{add}),
\label{eq:llm_add}
\end{equation}
where $\{(e_i, a_i)\}^L_{i=1}$ denotes a collection of entity-attitude pairs of length $L$ with $e_i$ being the entity and $a_i$ being the corresponding attitude. For the sake of brevity, we omit the post-processing step on the text generated by LLMs in Eq~\eqref{eq:llm_add} and only display the final results $\{(e_i, a_i)\}^L_{i=1}$. Specifically, $P_{add}$ is the prompt template for add operation with task descriptions such as \textit{``Given a conversation, summarize multiple entities and the user's attitudes towards these entities. Entities include movie titles and related attribute features.''} and format requirements like \textit{``The output format should be in JSON, with the entity as the key and the attitude as the value.''}  Simultaneously, we record a timestamp $ts_{gen}$ for this generation to facilitate subsequent deletion operations. Subsequently, we add each entity $e_i$ as a key, and its corresponding attitude $a_i$ and timestamp as a value to  $\mathcal{M}_u$.  For each entity-attitude pairs $(e_i, a_i)$, we have
\begin{equation}
   \mathcal{M}_u[e_i] \leftarrow \text{Write}(a_i,ts_{gen}),
\label{eq:llm_write}
\end{equation}
where $\text{Write}(\cdot)$ denotes the write operation to memory bank $\mathcal{M}_u$. 

    

\textbf{Merge} operation comes into play when adding the entity $e_i$ extracted from dialogue and countering a conflict; that is, the extracted entity already exists in the memory bank $\mathcal{M}_u$. In such cases, we need to merge the newly generated attitude $a_i$ with the existing attitude $\hat{a_i}$. The LLMs also carry out this merge operation in Eq~\eqref{eq:llm_merge}, encapsulating $a_i$ and $\hat{a_i}$ into prompt template $P_{merge}$,
\begin{equation}
    a_i^* = f_{LLM}(a_i, \hat{a_i}, P_{merge}),
\label{eq:llm_merge}
\end{equation}
where $a_i^*$ denotes the merged attitude generated by LLMs and the prompt template $P_{merge}$ contains task description such as, \textit{``Given a user's existing and new attitudes, merge these two attitudes, prioritizing the new attitude in case of conflicts.''} Then the merged attitude $a_i^*$ is inserted into the memory bank $\mathcal{M}_u$, with the timestamp $ts_{merge}$ updated upon the completion of this operation,
\begin{equation}
   \mathcal{M}_u[e_i] \leftarrow \text{Write}(a_i^*,ts_{merge}).
\label{eq:llm_write}
\end{equation}
Furthermore, the merge operation can also be extended from the same entity to several semantically similar entities, such as "phone" and "mobile phone", which further improves storage efficiency.


    


\textbf{Delete} operation is rule-based and does not require the intervention of LLMs. Additionally, it is considered as an optional procedure. In cases where storage capacity is constrained, we set a time period threshold $D$ and conduct periodic scanning over the entire memory bank. We then delete entities that have not been operated on (\ie, add, merge, or retrieve) for at least a period of time $D$ according to the recorded timestamps. This process is outlined as follows:
\begin{equation}
   \mathcal{M}_u \leftarrow \text{Delete}(\mathcal{M}_u, D).
\label{eq:llm_delete}
\end{equation}
where $\text{Delete}(\cdot)$ represents deleting each record in $\mathcal{M}_u$ whose timestamp has a time difference exceeding $D$ with the current time. Note that this is just one implementation method for the delete operation. Other strategies, such as usage frequency, can also be considered.


\subsubsection{Memory Retrieval}\label{sec:retrieval}
Now that we have established a personalized memory bank for user $u$ from his/her historical dialogue sessions $H_u$, not all memories within it are necessarily conducive to his/her current dialogue session $C_T$'s $k$-th turn. 
Utilizing all memories will introduce noise and possibly exceed the context window of LLMs, leading to inferior performance. Therefore, it becomes imperative to retrieve relevant memories from the memory bank based on the current dialogue session. Previous works on memory retrieval mostly use vector similarity retrieval methods like cosine similarity~\cite{zhong2024memorybank,li2023mot},  which can quickly process large amounts of data but may also retrieve some unrelated content. LLMs perform better in judging relevance~\cite{abbasiantaeb2024can,upadhyay2024llms} but have difficulty handling large candidate sets. Thus, we combine the two approaches -- first, adopt vector similarity to obtain candidate entities and then leverage LLMs to further discern relevant entities.


Specifically, we first utilize vector similarity retrieval methods, \eg, cosine similarity, as a preliminary filtration step~\cite{zhong2024memorybank,li2023mot} to obtain a candidate entity list $\hat{\mathcal{E}_u}$. When the number of entities in the memory bank is relatively limited, we can omit this step and directly use all the entities of $\mathcal{M}_u$ as $\hat{\mathcal{E}_u}$. Next, we incorporate $\hat{\mathcal{E}_u}$ and the previous $k-1$ rounds of utterances $\{s_j\}^{k-1}_{j=1}$ of the current session into the prompt template $P_{retrieve}$ in Eq~\eqref{eq:llm_retrieve}, allowing LLMs to select relevant entities.
\begin{equation}
    \mathcal{E}_u = f_{LLM}(\hat{\mathcal{E}_u}, \{s_j\}^{k-1}_{j=1}, P_{retrieve}),
\label{eq:llm_retrieve}
\end{equation}
where $\mathcal{E}_u$ is the retrieved entity list relevant to the ongoing conversation and $P_{retrieve}$ includes task description like \textit{``Select the $Q$ most relevant entities from the entity list based on the user's needs in the conversation, sorted by relevance''}, with output format being list and $Q$ is the hyperparameter. With $\mathcal{E}_u$, we can obtain the corresponding attitude list $\mathcal{A}_u$
  for the user from $\mathcal{M}_u$ as demonstrated in Eq~\eqref{eq:llm_read}. This refined and relevant information will assist the LLMs' recommendations in the subsequent Section~\ref{sec:integration}.
\begin{equation}
   \mathcal{A}_u = \text{Read}(\mathcal{M}_u, \mathcal{E}_u),
\label{eq:llm_read}
\end{equation}
where $\text{Read}(\cdot)$ means reading corresponding attitude from $\mathcal{M}_u$ for each entity in $\mathcal{E}_u$. Note that the read operation also modifies the timestamp of the corresponding entity.

    
    
    
  
In user-specific memory, the frequency of invoking LLMs is not high. This stems from the limited number of memory update operations. LLMs are usually only called once for an add operation at the end of each dialogue session, and the merge operation is rarely needed except in cases of minimal entity overlap. Not to mention that the delete operation does not involve LLMs at all. Moreover, the memory retrieval operation is exclusively activated upon recommendation, and LLMs are called once on the pre-filtered candidate entities for each retrieval.

\subsection{General Memory (GM)}\label{sec:UM}
Although we have derived user-specific memory from the user's historical dialogues, possessing such memory alone is insufficient for conversational recommendations. On the one hand, previous conversational agents or assistants~\cite{zhong2024memorybank,liu2023think,park2023generative,lu2023memochat} typically only consider user-specific memory unique to each user because they are not involved in the recommendation task inherent to conversational recommendations. This task relies not only on a user's individual preferences but also on collaborative knowledge, which involves inferring a user's preferences based on the preferences of similar users. Such insights cannot be derived from a single user's historical dialogues; instead, the model needs a comprehensive understanding of the overall data distribution.
On the other hand, not all users possess sufficient historical dialogue sessions to form enough user-specific memories. This raises the issue of cold-start users with limited memory. Enhancing the performance on these cold-start users is also a critical concern that warrants consideration.

To this end, beyond user-specific memory, we also need to consider some communal knowledge shared among users, referred to as general memory. Here, we primarily focus on two aspects: collaborative knowledge and LLM-driven reasoning guidelines. Training LLMs to acquire embedded collaborative knowledge incurs significant costs. Hence, we integrate an external, specialized expert model dedicated to extracting collaborative signals, providing LLMs with low-cost collaborative knowledge. The experiential insights derived from the reasoning of LLMs also constitute a crucial part of knowledge shared by users. Thus, we maintain a repository to house the reasoning knowledge and experience LLMs acquire during their reasoning process. Both categories of knowledge are concise and readily usable without retrieval. Their combined effect can enhance the efficacy of recommendations, particularly benefiting recommendations for cold-start users.
\subsubsection{Collaborative Knowledge}
A pivotal element in recommendation tasks is collaborative knowledge, which encapsulates shared patterns extracted from a myriad of user behaviors.  It facilitates recommender systems in uncovering commonalities and trends among user groups, ensuring precise recommendations. Moreover, it aids in providing tailored recommendations to new users by analyzing the preferences and behaviors of analogous user clusters, effectively mitigating the issue of cold-start users. Collaborative knowledge often requires the model to understand the overall data distribution, typically achieved by training on the entire dataset. Given the substantial costs of training LLMs, we adopt external specialized models for extracting this knowledge, thereby endowing LLMs with cost-effective collaborative insights.

Specifically, for the $k$-th turn of the current session $C_T$, the input to the expert model $g(\cdot)$ comprises the preceding $k-1$ turns of utterances $\{s_j\}^{k-1}_{j=1}$, and the items mentioned in these turns $\{\mathcal{I}_j\}^{k-1}_{j=1}$, where $\mathcal{I}_j$ represents the set of items mentioned by the user or the recommender in the $j$-th turn of utterance.  The expert model's prediction of the recommendations $\hat{\mathcal{I}_k}$ is derived as follows:
\begin{equation}
   \hat{\mathcal{I}_k} = g(\{s_j\}^{k-1}_{j=1}, \{\mathcal{I}_j\}^{k-1}_{j=1}).
\label{eq:expert_model}
\end{equation}

\subsubsection{Reasoning Guidelines}
Given that we employ LLMs for recommendations, the reasoning guidelines of LLMs are yet another pivotal knowledge that should be shared among users. Previous studies have found that extracting natural language experience from various decision-making tasks can be beneficial for subsequent tasks~\cite{zhao2024expel,zhang2024large}.  Consequently, we continuously prompt LLMs to reflect on the current reasoning process, extract experience from successful or failed examples, and use them to aid in subsequent reasoning tasks. Specifically, we begin by providing a manually crafted set of simple reasoning guidelines  $\mathcal{R}$ as an initialization. This set includes basic reasoning rules such as \textit{"Let's think step by step"} and \textit{"Consider user's needs during conversations"}. Subsequently, we integrate LLMs' reasoning trajectory $t$ and outcomes of recommendation $o$ into a dynamic prompt template $P_{reflect}$ in Eq~\eqref{eq:llm_reason}, allowing for iterative updates to the evolving reasoning guideline set $\mathcal{R}$ based on LLMs' learning and experience.
\begin{equation}
    \mathcal{R} \leftarrow f_{LLM}(t, o, \mathcal{R}, P_{reflect}),
\label{eq:llm_reason}
\end{equation}
where $t$ denotes LLMs' reasoning trajectory on the final recommendation in Section~\ref{sec:integration}, encapsulating the original prompt and the step-by-step reasoning process, and $o$ represents the user's text response to the recommendation, indicating whether the user is satisfied with the recommendation. The prompt template $P_{reflect}$ is equipped with the task description like \textit{``Given the reasoning process and its result, summarize the experience for successful reasoning and reflect on the experience for the failure of unsuccessful reasoning. Then, update the current reasoning guideline set and keep the total number of experiences within 10.''}

    
    
    
    

Collaborative knowledge extraction is conducted through an expert model, independent of LLMs involvement. The extraction of reasoning guidelines does not necessitate frequent updates and can occur at more extended intervals. Consequently, the utilization of general memory does not lead to frequent invocations of LLMs.

\subsection{Integration with Memory for CRSs}\label{sec:integration}
After introducing user-specific memory and general memory, we will integrate the two kinds of knowledge into the prompt to enable LLMs to generate recommendations during conversations. For the $k$-th turn of current session $C_T$, we retrieve relevant entity and attitude lists $\mathcal{E}_u$ and $\mathcal{A}_u$ from user $u$'s personalized memory bank $\mathcal{M}_u$ based on the previous $k-1$ rounds of utterances $\{s_j\}^{k-1}_{j=1}$, as described in Section~\ref{sec:retrieval}. Next, we obtain collaborative knowledge from the expert model using Eq.~\eqref{eq:expert_model}, \ie, the predicted recommendation list $\hat{\mathcal{I}_k}$
 . These pieces of information, along with reasoning guideline set $\mathcal{R}$ and the previous utterances $\{s_j\}^{k-1}_{j=1}$, are combined in prompt $P_{rec}$ and fed into LLMs to get recommendations, 

\begin{equation}
    \widetilde{\mathcal{I}_k} = f_{LLM}\big(\{s_j\}^{k-1}_{j=1}, \mathcal{E}_u, \mathcal{A}_u, \hat{\mathcal{I}_k}, \mathcal{R}, P_{rec}\big),
\label{eq:llm_rec}
\end{equation}
where $\widetilde{\mathcal{I}_k}$ denotes the recommended item list generated by LLMs. The prompt template $P_{rec}$ can leverage the ability of both expert model and LLMs, with the task description such as \textit{``Based on the user's conversation, reasoning guidelines, and historical memory, select the top 20 movies from the expert model's recommended movie list that best fit the user's needs. If there are fewer than 20 movies, supplement them with relevant movies based on your own knowledge.''} 

    
    
    

    
\begin{table*}[]
\centering
\vspace{-5pt}
\caption{Comparison of different models on recommendation task. The best result is given in bold, while the second-best value is underlined. The symbol * indicates statistically significant improvement over the best baseline with $p < 0.01$.}
\vspace{-5pt}
\scalebox{0.81}{
\setlength{\tabcolsep}{0.5mm}{
\begin{tabular}{cccc|ccc|ccc|ccc|ccc|ccc}
\toprule
\multirow{3}{*}{Model} & \multicolumn{9}{c|}{TGReDial} & \multicolumn{9}{c}{ReDial} \\
\cmidrule{2-19}
 & \multicolumn{3}{c|}{HR} & \multicolumn{3}{c|}{MRR} & \multicolumn{3}{c|}{NDCG} & \multicolumn{3}{c|}{HR} & \multicolumn{3}{c|}{MRR} & \multicolumn{3}{c}{NDCG} \\
 \cmidrule{2-19}
 & @5 & @10 & @20 & @5 & @10 & @20 & @5 & @10 & @20 & @5 & @10 & @20 & @5 & @10 & @20 & @5 & @10 & @20 \\
 \midrule
ReDial & 0.0030 & 0.0055 & 0.0102 & 0.0015 & 0.0018 & 0.0021 & 0.0018 & 0.0027 & 0.0038 & 0.0293 & 0.0413 & 0.0882 & 0.0174 & 0.0190 & 0.0223 & 0.0203 & 0.0242 & 0.0361 \\
KBRD & 0.0050 & 0.0112 & 0.0174 & 0.0026 & 0.0035 & 0.0040 & 0.0032 & 0.0053 & 0.0069 & 0.0824 & 0.1395 & 0.2185 & 0.0337 & 0.0418 & 0.0470 & 0.0457 & 0.0646 & 0.0842 \\
KGSF & \underline{0.0112} & 0.0149 & 0.0249 & 0.0039 & 0.0044 & 0.0051 & 0.0057 & 0.0068 & 0.0094 & 0.0908 & 0.1378 & 0.2319 & 0.0385 & 0.0445 & 0.0508 & 0.0514 & 0.0663 & 0.0898 \\
TGReDial & 0.0075 & \underline{0.0174} & 0.0236 & 0.0055 & 0.0068 & 0.0072 & 0.0059 & 0.0091 & 0.0107 & 0.0874 & 0.1395 & 0.2336 & 0.0433 & 0.0502 & 0.0567 & 0.0543 & 0.0710 & 0.0948 \\
UCCR & 0.0087 & \underline{0.0174} & \underline{0.0286} & \underline{0.0071} & \underline{0.0082} & \underline{0.0090} & \underline{0.0075} & \underline{0.0103} & \underline {0.0131} & 0.1059 & 0.1782 & \underline{0.2672} & 0.0443 & 0.0538 & 0.0596 & 0.0594 & 0.0826 & \underline{0.1047} \\
UniCRS & 0.0050 & 0.0124 & 0.0236 & 0.0024 & 0.0035 & 0.0042 & 0.0031 & 0.0055 & 0.0083 & 0.1005 & 0.1605 & 0.2480 & 0.0392 & 0.0467 & 0.0528 & 0.0542 & 0.0731 & 0.0952 \\
ZSCRS & 0.0025 & 0.0087 & 0.0100 & 0.0007 & 0.0016 & 0.0017 & 0.0012 & 0.0032 & 0.0035 & \underline{0.1261} & \underline{0.1882} & 0.2353 & \underline{0.0511} & \underline{0.0594} & \underline{0.0629} & \underline{0.0695} & \underline{0.0896} & 0.1018 \\
\textbf{MemoCRS} & \textbf{0.0162*} & \textbf{0.0261*} & \textbf{0.0323*} & \textbf{0.0095*} & \textbf{0.0108*} & \textbf{0.0112*} & \textbf{0.0111*} & \textbf{0.0143*} & \textbf{0.0158*} & \textbf{0.1361*} & \textbf{0.2151*} & \textbf{0.2857*} & \textbf{0.0718*} & \textbf{0.0821*} & \textbf{0.0871*} & \textbf{0.0875*} & \textbf{0.1128*} & \textbf{0.1308*}
 \\
\bottomrule
\end{tabular}
}
}
\vspace{-5pt}
\label{tab:overall_rec}
\end{table*}
\section{Experiments} 
To gain more insights into MemoCRS, we tend to address the following research questions (RQs) in this section. 
\begin{itemize}
    \item \textbf{RQ1:} How does MemoCRS perform in recommendation and dialogue generation tasks in sequential CRSs? 
    \item \textbf{RQ2:}  What roles do the various modules of MemoCRS play in its performance? 
    \item \textbf{RQ3:} How effective and efficient is the user-specific memory?
    \item \textbf{RQ4:}  Can general memory address the issue of cold-start users?
\end{itemize}
 
\vspace{-4pt}
\subsection{Experiment Setups}
\subsubsection{Dataset}
We conduct experiments on two public datasets, a Chinese dataset TGReDial\footnote{\url{https://github.com/RUCAIBox/TG-ReDial}} and an English dataset ReDial\footnote{\url{https://redialdata.github.io/website/}}. \textbf{TGReDial}~\cite{tgredial} is a collection of Chinese conversational
recommendation sessions constructed in a semi-automatic topic-guided way. It contains 10,000 sessions of 129,392 utterances involving 1,482 users and 33,834 movies. \textbf{ReDial}~\cite{redial} is an English conversational recommendation dataset built manually by constructed through crowd-sourcing workers on
Amazon Mechanical Turk (AMT). It includes 10,006 dialogues of 182,150 utterances related to 51,699 movies and 504 users. 
As we emphasize the crucial role of preferences within historical dialogue sessions, we mainly follow the approach outlined in~\cite{uccr} to resplit the two datasets into train/valid/test sets based on \textbf{chronological order}. We randomly select a subset of users, with their last several sessions as the valid and test sets and the remaining sessions as the training set. Previous works find that repeated items can create shortcuts~\cite{zesrec}. Therefore, we also filter out conversations with duplicate items, ensuring that recommended items are not mentioned in previous conversation turns. Notably, some users in Redial lack dialogue sessions in the training set, indicating they have no historical dialogue sessions. These users are utilized to evaluate MemoCRS's performance on cold-start users.

\subsubsection{Baselines} To validate the effectiveness of MemoCRS, we select several representative methods in CRSs as our baselines. \textbf{ReDial}~\cite{redial} adopts an auto-encoder as the recommender and HRED~\cite{sordoni2015hierarchical} for dialogue generation. \textbf{KBRD}~\cite{kbrd} uses external knowledge graph DBPedia~\cite{bizer2009dbpedia} for the entities in dialogues to enhance the model's performance. \textbf{KGSF}~\cite{kgsf} incorporates two KGs, ConceptNet~\cite{speer2017conceptnet} and DBPedia~\cite{bizer2009dbpedia} to enhance the representations of words and entities, and uses Mutual Information Maximization to align them. \textbf{TGReDial}~\cite{tgredial} is proposed with the TGReDial dataset and incorporates a topic prediction task to enhance performance. \textbf{UniCRS}~\cite{UniCRS} unifies the recommendation and conversation tasks into the prompt learning paradigm, and utilizes fixed PLMs to fulfill both tasks in a unified approach. \textbf{UCCR}~\cite{uccr} jointly models current dialogue sessions, historical dialogue sessions, and look-alike users via a user-centric manner. \textbf{ZSCRS}~\cite{zesrec,wang2023rethinking} employs LLMs as zero-shot conversational recommenders. 
Here, for a fair comparison, we employ GPT4 (gpt-4-1106-preview) as its backbone LLM.

\subsubsection{Evaluation Metrics}In CRSs' evaluation, there are typically two tasks: recommendation and dialogue generation. However, LLMs often exhibit stronger dialogue generation capabilities than the smaller PLMs used in previous methods. Therefore, the comparison often focuses solely on their recommendation abilities in previous works that use LLMs as zero-shot recommenders~\cite{wang2023rethinking,zesrec}. In this work, our main emphasis is also on recommendation performance, but we also touch upon the task of dialogue generation.
For recommendation, several widely used metrics, \textit{HR@K}, \textit{NDCG@K}~\cite{ndcg}, and MRR, are adopted, following previous works~\cite{uccr,kbrd,UniCRS,kgsf}. Here, $K\in\{5,10,20\}$. 
For dialogue generation, we adopt human evaluations, where three annotators score the \textit{Fluency} and \textit{Informativeness} of the generated responses following~\cite{UniCRS,uccr}. The range of scores is from 0 to 2, and the scores of three annotators are averaged.
\subsubsection{Implementation Details}
Apart from ZSCRS, all other baselines are implemented with the open-source toolkit CRSLab~\cite{crslab}, and we conduct careful hyperparameter tuning to achieve the best performance. Data preprocessing also slightly differs from the original CRSLab; we change the dataset splitting method from random to chronological, using the users' last few sessions as the validation and test sets following~\cite{uccr}. Both ZSCRS and MemoCRS utilize GPT-4 (gpt-4-1106-preview) as the LLM backbone for generating recommendations, with a temperature parameter of 0. Their experiments are also conducted based on newly generated data from CRSLab. For MemoCRS\footnote{Our code will be available at \url{https://github.com/mindspore-lab/models/tree/master/research/huawei-noah/memocrs}}, we choose UCCR as the expert model. The number of retrieved memories $Q$ used in the final recommendation is 3 for TGReDial and 1 for ReDial. The number of candidates, \ie, $|\hat{\mathcal{I}_k}|$, provided by the expert model is 40, and the length of the final output recommendation list generated by LLMs is set at 20. 

\subsection{Effectiveness Comparison (RQ1)}
\subsubsection{Recommendation Task}

To validate the effectiveness of our proposed MemoCRS on recommendation task, we compare it with selected state-of-the-art baselines in CRSs. The results are presented in Table~\ref{tab:overall_rec}, from which we have the following observations:
\begin{itemize}
    \item Our proposed MemoCRS significantly outperforms the baselines. For instance, on the TGReDial dataset, MemoCRS shows improvements of 13.04\% in HR@20, 23.73\% in MRR@20, and 20.41\% in NDCG@20 over the strongest baseline. On ReDial, these enhancements are 6.93\%, 38.49\%, and 24.96\%, respectively. This indicates the effectiveness of incorporating memory-enhanced LLMs and modeling the continuity of user preferences in CRSs.

    \item Methods that utilize user history generally outperform those that do not. Models like TGRedial, which leveraged previously interacted items, and URCC, which uses historical dialogue sessions, achieve better results than other baselines. MemoCRS, which incorporates memory techniques to refine user historical preferences, yields superior outcomes and validates the importance of modeling user preference continuity and sequentiality in CRSs.

    \item Leveraging larger PLMs often leads to better performance, albeit influenced by the dataset. 
    By incorporating LLMs, simple zero-shot prompting methods (\eg, ZRCRS) are able to outperform carefully designed and fine-tuned small PLMs (\eg, UCCR and UniCRS) on ReDial, which has been corroborated in previous studies as well~\cite{zesrec,wang2023rethinking}.
    However, this aspect is also subject to dataset influences. On TGReDial, the performance of zero-shot prompting was notably poor, possibly due to LLMs (GPT-4) having a limited understanding of Chinese and Chinese movies, coupled with the dataset primarily featuring niche movies.
\end{itemize}

\begin{table}[h]
\centering
\caption{Comparison on dialogue generation task.}
\vspace{-5pt}
\scalebox{0.82}{
\setlength{\tabcolsep}{2mm}{
\begin{tabular}{ccc|cc}
\toprule
\multirow{2}{*}{Model} & \multicolumn{2}{c|}{TGReDial} & \multicolumn{2}{c}{ReDial} \\
\cmidrule{2-5}
 & Fluency & Informativeness & Fluency & Informativeness \\
 \midrule
ReDial & 0.45 & 0.40 & 0.41 & 0.40 \\
KBRD & 1.04 & 0.99 & 0.97 & 1.01 \\
KGSF & 1.04 & 1.08 & 1.13 & 1.15 \\
TGReDial & 0.80 & 0.84 & 0.88 & 0.87 \\
UCCR & \underline{1.10} & \underline{1.12} & 1.11 & 1.17 \\
UniCRS & 0.42 & 0.43 & \underline{1.18} & \underline{1.19} \\
LLM & \textbf{1.87*} & \textbf{1.82*} & \textbf{1.86*} & \textbf{1.87*} \\
\bottomrule
\end{tabular}
}}
\vspace{-5pt}

\label{tab:overall_gen}
\end{table}

\subsubsection{Dialogue Generation Task} 

Apart from the recommendation task, dialogue generation is also a crucial aspect of conversational recommendation. Considering that LLMs generally exhibit much better language generation capabilities compared to smaller PLMs used in previous works in CRSs~\cite{uccr,UniCRS,kbrd,kgsf}, most works that leverage LLMs for dialogue recommendation only focus on the recommendation task~\cite{zesrec,wang2023rethinking}. Some researchers found that LLMs excel in providing explainable recommendations and creating an interactive user experience~\cite{sun2024large}. Here, we quantitatively compare the quality of dialogues generated by LLMs (gpt-4-1106-preview) using zero-shot prompting with traditional CRSs through human evaluation. The results from Table~\ref{tab:overall_gen} demonstrate that the dialogues generated by LLMs are significantly more fluent and informative than those generated by traditional models in CRSs, and utilizing LLMs for CRSs can produce more natural and accurate responses. One important point to note is that UniCRS yields unsatisfactory results on the Chinese dataset TGReDial because it is built upon DialoGPT, which does not support Chinese well. 

\subsection{Ablation Study (RQ2)}\label{sec:ablation}
To investigate the impact of each component in MemoCRS, we design several variants in Table~\ref{tab:ablation}. The model variants \textbf{w/o UM}, \textbf{w/o CK}, and \textbf{w/o RG} represent removing user-specific memory, collaborative knowledge, and reasoning guidelines from MemoCRS, respectively. \textbf{Manual RG} replaces the reasoning guidelines summarized by LLMs with the manually initialized reasoning guidelines. 
From the results in Table~\ref{tab:ablation}, we can observe that after removing collaborative knowledge, reasoning guidelines, and user-specific memory, the model's performance shows a noticeable decline, indicating that both the user-specific memory and general memory we designed have significant impacts on the model's performance. Additionally, when replacing the reasoning guidelines summarized by LLMs with manually crafted guidelines, there is also a decrease in performance, suggesting that reasoning guidelines generated by LLMs complement missing parts in human summarization and are more suitable for guiding LLMs' reasoning. 

\begin{table*}[h]
\centering
\caption{Ablation study of MemoCRS. The best result is given in bold, while the second-best value is underlined.}
\vspace{-4pt}
\scalebox{0.8}{
\setlength{\tabcolsep}{0.8mm}{
\begin{tabular}{cccc|ccc|ccc|ccc|ccc|ccc}
\toprule
\multirow{3}{*}{Variants} & \multicolumn{9}{c|}{TGReDial} & \multicolumn{9}{c}{ReDial} \\
\cmidrule{2-19}
 & \multicolumn{3}{c|}{HR} & \multicolumn{3}{c|}{MRR} & \multicolumn{3}{c|}{NDCG} & \multicolumn{3}{c|}{HR} & \multicolumn{3}{c|}{MRR} & \multicolumn{3}{c}{NDCG} \\
 \cmidrule{2-19}
  & @5 & @10 & @20 & @5 & @10 & @20 & @5 & @10 & @20 & @5 & @10 & @20 & @5 & @10 & @20 & @5 & @10 & @20 \\
\textbf{MemoCRS} & \textbf{0.0162} & \textbf{0.0261} & \textbf{0.0323} & \textbf{0.0095} & \textbf{0.0108} & \textbf{0.0112} & \textbf{0.0111} & \textbf{0.0143} & \textbf{0.0158} & \textbf{0.1361} & \textbf{0.2151} & {\underline{0.2857}} & \textbf{0.0718} & \textbf{0.0821} & \textbf{0.0871} & \textbf{0.0875} & \textbf{0.1128} & \textbf{0.1308} \\
w/o UM & \underline{0.0149} & 0.0211 & 0.0299 & \underline{0.0071} & \underline{0.0078} & \underline{0.0084} & \underline{0.0090} & \underline{0.0109} & \underline{0.0131} & \underline{0.1160} & 0.1966 & 0.2807 & \underline{0.0623} & \underline{0.0731} & \underline{0.0789} & \underline{0.0755} & \underline{0.1017} & \underline{0.1228} \\
w/o CK & 0.0087 & 0.0100 & 0.0124 & 0.0060 & 0.0061 & 0.0063 & 0.0066 & 0.0070 & 0.0076 & \underline{0.1160} & 0.1748 & 0.2319 & 0.0615 & 0.0692 & 0.0730 & 0.0750 & 0.0938 & 0.1072 \\
w/o RG & 0.0137 & \underline{0.0224} & 0.0274 & 0.0055 & 0.0065 & 0.0068 & 0.0075 & 0.0102 & 0.0114 & 0.1025 & 0.1933 & \textbf{0.2924} & 0.0483 & 0.0604 & 0.0676 & 0.0615 & 0.0909 & 0.1164 \\
Manual RG & 0.0112 & 0.0211 & \underline{0.0311} & 0.0054 & 0.0068 & 0.0075 & 0.0069 & 0.0101 & 0.0126 & 0.0958 & \underline{0.2050} & 0.2807 & 0.0499 & 0.0649 & 0.0706 & 0.0611 & 0.0969 & 0.1176 \\
\bottomrule
\end{tabular}
}}
\vspace{-5pt}
\label{tab:ablation}
\end{table*}

\subsection{Analysis of User-Specific Memory}\label{sec:memory}
\subsubsection{Memory Efficiency}
This section delves into the memory efficiency of user-specific memory. Due to the compact nature of collaborative knowledge and reasoning guidelines utilized without retrieval, general memory is inherently efficient. On the other hand, user-specific memory compresses the information from a user's historical dialogue sessions. Hence, we focus on the efficiency of user-specific memory. Table~\ref{tab:avg_token} presents the average number of tokens for each user under different memory scenarios. \textbf{"Total Dialogues"} refers to simply including all the user's historical sessions without refinement and retrieval as memory. \textbf{"Total UM"} indicates using all user-specific memory refined by our proposed MemoCRS without retrieval. \textbf{"Our"} represents the retrieved user-specific memory in our MemoCRS that is relevant to the current conversation.

From Table~\ref{tab:avg_token}, we can observe that the token count of all user-specific memory refined by MemoCRS is much lower than that of all historical dialogues on both datasets. 
On ReDial, the difference is even up to 4.6 times, indicating the significant redundancy in the original historical dialogues, and the entity-based memory bank in MemoCRS can reduce this redundancy. Furthermore, retrieving relevant memory with LLMs significantly reduces the token consumption, reducing the input token count for LLMs and thus lowering the inference cost. This also suggests that there is not a large number of relevant memories in the memory bank, so we need to retrieve and extract relevant information to eliminate noise.

\begin{table}[h]
\centering
\caption{Average number of tokens pre user for different types of memory. UM denotes user-specific memory.}
\vspace{-5pt}
\scalebox{0.9}{
\setlength{\tabcolsep}{1mm}{
\begin{tabular}{ccc}
\toprule
Avg. \#Token Per User & TGReDial & ReDial \\
\midrule
Total Dialogues & 1781.89 & 7154.07 \\
Total UM & 1026.98 & 1526.51 \\
Ours & 21.41 & 50.27 \\
\bottomrule
\end{tabular}
}
}
\label{tab:avg_token}
\vspace{-10pt}
\end{table}
\subsubsection{Memory Effectiveness}
\begin{figure}[h]
    \centering
\vspace{-10pt}
    \includegraphics[width=0.43\textwidth]{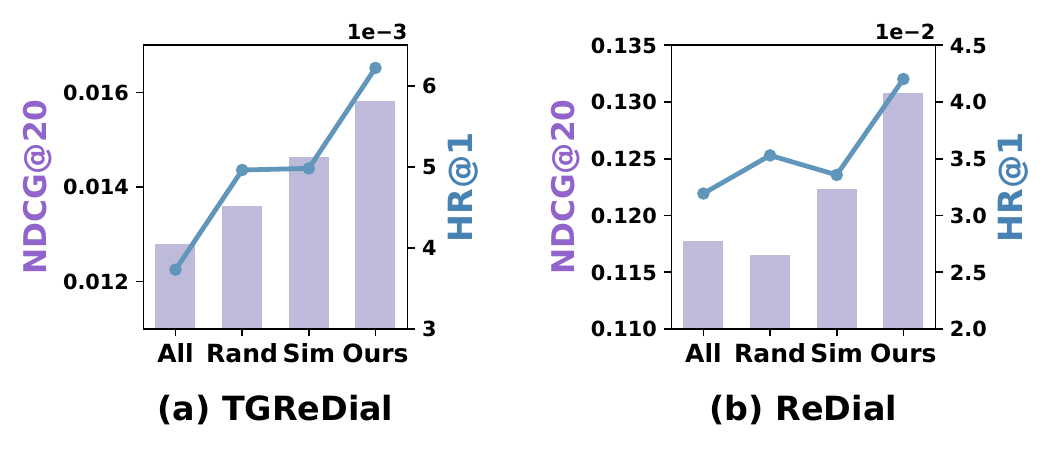}
\vspace{-10pt}
    \caption{Comparison between different kinds of memory.}
\vspace{-10pt}
    \label{fig:memo_effect}
\end{figure}

Next, we analyze the efficacy of user-specific memory design and its retrieval strategies. To this end, we devise several variants with different memory utilization approaches, based on the user-specific memory generated by MemoCRS.
\textbf{"All"} denotes the direct utilization of all refined user-specific memories without any retrieval operation. \textbf{"Rand"} represents randomly selecting $Q$ entity-attitude pairs from the user-specific memory bank to exclude the effect of the input length. \textbf{"Sim"} refers to encoding user-specific memory through BERT and employing cosine similarity to retrieve the most relevant $Q$ pairs. Lastly, \textbf{"Ours"} signifies our proposed two-stage retrieval, which first adopts cosine
similarity to obtain candidate entities and then leverages LLMs to further select the most relevant $Q$ pairs. Notably, the number of final entity-attitude pairs $Q$ used in the latter three variants is consistent across the same dataset and aligned with the experiments outlined in Table~\ref{tab:overall_rec}. We do not compare inputting all historical dialogue sessions here because their length may exceed the context window of LLMs, and we have already compared the model, \ie, UCCR, equipped with historical sessions in Table~\ref{tab:overall_rec}. We select two metrics, \textit{HR@1} and \textit{NDCG@20}, and plot the performance of different variants on two datasets in Figure~\ref{fig:memo_effect}.

From Figure~\ref{fig:memo_effect}, we can observe that using all historical memory yields the poorest results, indicating significant noise and irrelevant memory in the memory bank for the current conversation session. This underscores the necessity of retrieval mechanisms. Randomly selecting memories exhibits marginally better performance, suggesting that an abundance of tokens across all memories may also impact LLMs' decision-making process. Utilizing cosine similarity for retrieval leads to further enhancements, particularly in \textit{NDCG@20}, signifying its capability to extract more relevant memories. However, there is a slight decrease in effectiveness in \textit{HR@1}. Our approach of leveraging LLMs for further memory extraction showcases superior efficacy, highlighting LLMs' ability to accurately assess the relevance of memory, thereby effectively extracting pertinent information while mitigating noise.

\subsection{Cold-Start Users (RQ4)}\label{sec:cold_start}
\begin{figure}[h]
    \centering
    \includegraphics[width=0.45\textwidth]{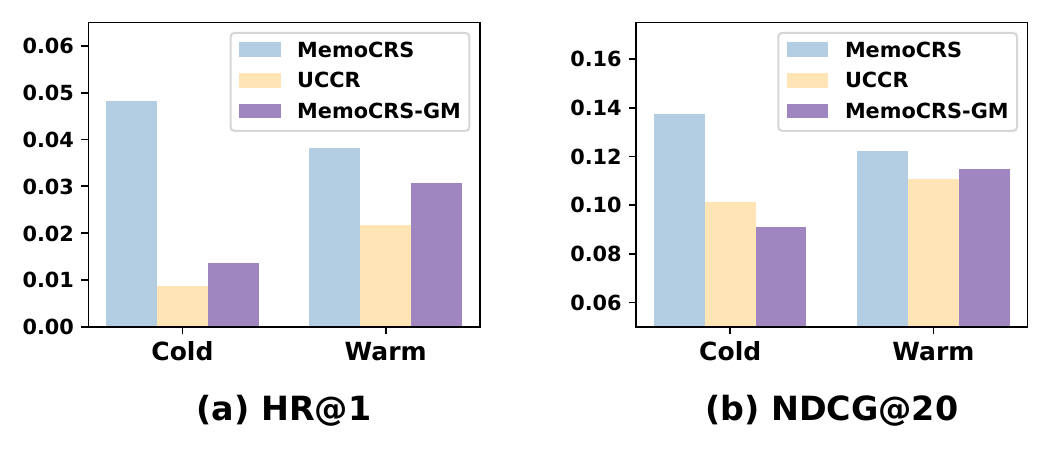}
    \caption{Performance comparison on cold and warm users.}
    \label{fig:cold_start}
\end{figure}

Given limited user-specific memory for cold-start users, we design a shared general memory among users to address this issue. In this section, we primarily investigate whether our proposed model, especially the general memory, can improve the recommendations for cold-start users. On the ReDial dataset, we divide the test set into \textbf{Warm} and \textbf{Cold} groups based on whether the user has historical sessions in the training set. Subsequently, we compare the performance of our proposed \textbf{MemoCRS}, the best baseline \textbf{UCCR}, and MemoCRS without general memory, \ie, \textbf{MemoCRS-GM}, on these two groups. We adopt \textit{HR@1} and \textit{NDCG@20} to assess both recall and ranking performance, with the results depicted in Figure~\ref{fig:cold_start}. 
From these results, we draw the following conclusions.

Firstly, MemoCRS exhibits notable enhancements over the best baseline UCCR on both warm and cold user groups, with a particularly significant uplift observed on cold-start users. While UCCR demonstrates superior performance in \textit{HR@1} and \textit{NDCG@20} for warm users, MemoCRS excels in enhancing the experience for cold users. This indicates that MemoCRS's design can benefit both cold and warm users, with a pronounced impact on cold-start users.
Secondly, MemoCRS without general memory (MemoCRS-GM) exhibits a decrease in performance compared to MemoCRS across two user groups, particularly pronounced on cold-start users. Although MemoCRS-GM still outperforms UCCR on warm users, its performance of \textit{HR@1} for cold-start users lags behind UCCR's. This suggests that general memory can aid both cold and warm user groups, with a more significant impact on cold-start users, demonstrating the effectiveness of general memory in mitigating cold-start issues.


\section{Conclusion}
In this work, we highlight the importance of user preference continuity for sequential CRSs and, for the first time, introduce memory-enhanced LLM to refine and manage user preference. We propose MemoCRS, which constitutes user-specific memory tailored for each user's personalized preferences and general memory containing universal knowledge shared across different users. On both Chinese and English datasets, MemoCRS shows superior performance compared to baselines.

\bibliographystyle{ACM-Reference-Format}
\bibliography{sample-base}


\end{document}